# Advances in Thick GEM-like gaseous electron multipliers. Part I: atmospheric pressure operation

C. Shalem, R. Chechik∗, A. Breskin and K. Michaeli

*Dept. of Particle Physics*
*The Weizmann Institute of Science, 76100 Rehovot, Israel.*

**Abstract**

Thick GEM-like (THGEM) gaseous electron multipliers are made of standard printed-circuit board perforated with sub-millimeter diameter holes, etched at their rims. Effective gas multiplication factors of $10^5$ and $10^7$ and fast pulses in the few nanosecond rise-time scale were reached in single- and cascaded double-THGEM elements, in atmospheric-pressure standard gas mixtures with single photoelectrons. High single-electron detection efficiency is obtained in photon detectors combining THGEMs and semitransparent UV-sensitive CsI photocathodes or reflective ones deposited on the top THGEM face; the latter benefits of a reduced sensitivity to ionizing background radiation. Stable operation was recorded with photoelectron fluxes exceeding MHz/mm$^2$. The properties and some potential applications of these simple and robust multipliers are discussed.

∗ Corresponding Author. Email: Rachel.Chechik@weizmann.ac.il; Tel: 972-8-9344966 Fax: 972-8-9342611

# 1. Introduction

We describe the operation mechanism and recent advances in Thick GEM-like (THGEM) electron multipliers, operating at atmospheric pressure. The THGEM [1] is a robust, simple to manufacture, high-gain gaseous electron multiplier. Its operation is based on gas multiplication within small, sub-millimeter to millimeter diameter holes, in a standard double-face Cu-clad printed circuit board.

Gas avalanche multiplication within small holes is attractive because the avalanche-confinement in the hole strongly reduces photon-mediated secondary effects. In addition, hole-multiplication provides true pixilated radiation localization. Hole-multiplication has been the subject of numerous studies in a large variety of applications. Among them: optical particle tracking by gas discharge in capillary plates and tubes [2]; Gamma radiation detection with small diameter lead-glass and other tube-like converters, followed by charge multiplication within the holes [3, 4]; proportional amplification in other structures like the Micro-Well [5] and the glass Capillary Plates (CP) [6,7] etc.

The most attractive and extensively studied hole-multiplier is the Gas Electron Multiplier (GEM) [8], made of 50-70-μm diameter holes chemically etched in a 50-μm thick metalized Kapton foil. An electric potential applied between the GEM electrodes creates a strong dipole electric field within the holes, responsible for an efficient focusing of ionization electrons into the holes and their multiplication by gas avalanche process therein. The GEM operates in a large variety of gases, including noble-gas mixtures, providing a gain of $\sim10^4$ in a single element and gains exceeding $10^6$ in a cascade of 3-4 elements [9,10]. The avalanche process is fast (typical rise-time of a few ns) and free of photon-mediated secondary effects, due to the optical opacity of the GEM electrodes. In addition to its use for particle tracking [11] and in Time Projection Chambers (TPC) [12], the GEM can also be efficiently coupled to gaseous or solid radiation converters, resulting in a large variety of radiation detectors developed for imaging of x-rays [13,14], neutrons [15] and UV-to visible light [16].

A more recent hole-multiplier derived from the GEM is the Micro-Hole & Strip Plate (MHSP) [17,18]; it provides electron multiplication in GEM-like holes followed by a second multiplication stage on thin anode strips patterned on the bottom of the same electrode. High gains are reached in a single MHSP element, even in noble gas mixtures [19]. Cascaded MHSP and GEM multipliers present high gains and significantly reduced yield of avalanche ions back-flowing to the first element in the cascade, with an important impact on the detector's properties [20]. The success of GEMs and glass Capillary Plates triggered the concept of a coarser structure, named by its authors the "optimized GEM", made by drilling millimetric holes in a 2mm thick Cu-plated G-10 printed-circuit board (PCB) [21,22]. These multipliers yielded gains of $10^4$ in Ar/isobutane (95:5) and in pure Xe; gains of $10^3$ were reached in pure Xe in combination with a CsI photocathode (PC).

Our THGEM described in this work, is fabricated in standard PCB technique; unlike the "optimized GEM", our concept combines in addition to hole drilling in a PCB also chemical etching of the rim around each hole (Fig.1). The latter was found essential for reducing considerably discharges at the hole's rim, resulting in higher permissible voltages and higher detector gains.

The THGEM is mechanically an expansion of the standard GEM, with its various dimensions being enlarged by factors ranging from 5 to 50. But though the geometrical dimensions are expanded by large factors, most parameters governing its operation, e.g. operation voltage, electric fields, electron diffusion, etc. do not scale accordingly. Therefore, the optimization of the THGEM parameters required a broad systematic study. In the previous [1,23] and in the present works, we have investigated a large variety of THGEM geometries over a broad pressure range (0.5-760 Torr); we will discuss in this and in a following article [24] the optimal geometry in terms of hole diameter, hole spacing and electrode thickness, for different applications at atmospheric and at low gas pressures.

The THGEM operation principle, shown in Fig.2, is similar to that of the standard GEM. Upon application of a voltage difference across the THGEM, a strong dipole field $E_{hole}$ is established within the holes. Electrons deposited by ionizing radiation in a conversion region above the THGEM, or produced on a solid radiation converter (e.g. a PC), are drifting towards the THGEM under the field $E_{drift}$ and are focused into the THGEM holes by the strong electric field inside the holes. The radiation converter can be a *semitransparent (ST)* one, placed above the THGEM, or a *reflective (Ref)* one, deposited on the THGEM top surface. We denote these *semitransparent* and *reflective* modes, respectively. With a *Ref* PC the most appropriate $E_{drift}$ value is 0, like in a GEM [25], as discussed below; with a *ST* one, some finite drift field is required (see section 3.3 below). The electrons are multiplied within the holes under the high electric field (~ 25-50 kV/cm, see below); depending on the size and direction of the field $E_{trans}$, a fraction of the resulting avalanche electrons are collected on the THGEM bottom electrode while the rest may be further transferred to a collecting anode or to a second, possibly similar, multiplier element. Each hole acts as an independent multiplier; the avalanche confinement within the holes has the advantage of reduced photon-mediated secondary effects; this leads to high-gain operation in a large variety of gases, including highly scintillating ones like pure $CF_4$. Photon detectors having a *Ref* PC deposited on the THGEM top face are particularly interesting: in this geometry the PC is totally concealed from avalanche-induced photons and therefore no photon-feedback effects are present. As the latter are a major performance-limiting mechanism of photon-imaging detectors [20], their suppression is of an advantage for conceiving high-efficiency photon detectors with sensitive PCs. The role of ion-induced secondary effects [20], their amplitude in THGEMs and conditions for their reduction will be discussed below.

In this article we will concentrate on the THGEM's operation mechanism and properties at atmospheric pressure. We shall present results demonstrating the role of each geometrical and operational parameter of the THGEM. The operation and properties of photon detectors with *ST* and *Ref* PCs and of soft x-ray detectors will be described. The interesting THGEM properties at low gas pressures, of which some preliminary results are given in [1], will be the subject of another article [24].

## 2. Methodology

The present study encompasses the production of THGEM electrodes, calculations of electric fields by the MAXWELL software package [26], simulations of electron transport by the GARFIELD software [27] and systematic measurements of various operation properties of the THGEM. For the sake of clarity the details of each of the measurements will be provided in the next section, together with the relevant results.

*2.1. THGEM production procedure*

The THGEM electrodes were produced in the Printed Circuit Board (PCB) industry [28], by a standard drilling and etching process, out of double-clad G-10 plates. Some of the electrodes were produced in KEVLAR, as discussed below. We used plates of a thickness t = 0.4 – 3.2 mm; the insulator was first drilled with a hexagonal pattern of holes of a diameter d = 0.3 – 2 mm and a pitch a = 0.7 – 4 mm; and then the copper was etched at a 0.1mm distance around the hole's rim (Fig.1). A large assortment of THGEM electrodes was produced by this very economic method; table 1 summarizes the various THGEM geometries studied in the present work.

*2.2. MAXWELL and GARFIELD simulations*

MAXWELL software was used to calculate the electric field maps (direction and values) in the vicinity of the THGEM electrode, within its holes and at its surface (fig 3). They were fed into the GARFIELD simulation package, providing the electron and ion paths, including diffusion, and the electron multiplication within the holes (fig. 2, 4). These tools allowed examining a large variety of electrode geometries, verifying our measurements in a given gas mixture and understanding the processes involved in the THGEM operation.

*2.3. Experimental techniques*

We have measured the following properties:
- **Electron Transfer efficiency (ETE)** – the probability to focus the electron from its creation point into a hole. For a gas-ionization electron the ETE includes only the transport of the electron. For a photoelectron emitted from a PC, the ETE includes also the extraction efficiency from the PC into the gas, which depends on the electric field on the PC surface [29]. The ETE is an important parameter for any hole-multiplication detector, affecting its operation; e.g. the detection efficiency of single-electron events or the energy resolution of charged particles or x-rays inducing ionization electrons in the conversion gap preceding the THGEM. The ETE depends on the detector's operation mode and conditions; it was measured in various gases as function of the THGEM operation voltage in single-multiplier geometry.

- **Effective gain ($G_{eff}$)** – the product of the absolute multiplication factor in the holes and the ETE. $G_{eff}$ was measured in different gases, in various THGEM geometries; it was assessed both in single- and double-element cascaded modes; in the latter, $G_{eff}$ represents the product of the absolute multiplication factors in both THGEMs, the ETE into both multiplier's holes and the electron extraction efficiency from the first element into the gap between them. (note $G_{eff}$ does not include the charge transfer efficiency to the readout anode, as common in the literature).
- **Counting-rate response** – the dependence of pulse-height on the event rate.
- **Ion back flow fraction (IBF)** – the fraction of ions created in the final avalanche that flow back and are collected at the PC (or penetrate the ionization region; Fig.2).

We have also measured the x-ray energy resolution and the pulse rise-time in some gases.

All measurements, except the x-ray energy resolution, were carried out with photoelectrons emitted from a CsI PC, irradiated with UV light from a continuous Ar(Hg) lamp or from a spontaneously discharging $H_2$ lamp. The experimental setups for the different measurements are described in paragraph 3. The PC was either a thin (30nm) *ST* film, vacuum deposited on a Quartz window, pre-coated with a very thin (2-3nm) under-layer of Cr, or a thick (300nm) *Ref* film, vacuum-deposited on the THGEM's top face. The *ST* mode with the PC placed a few mm above the multiplier, represents the operation mode of a THGEM coupled to any source of electrons located in the gap above it; besides the photomultiplier configuration, it could be a conversion gas gap for ionizing particles in a tracking detector or in a TPC, an x-ray conversion gap or another multiplier preceding the THGEM.

We used an individual power supply for each electrode, permitting to independently vary the different fields. A current limit of 50nA was usually set on the power supplies biasing the THGEMs (CAEN, model N471A) and a 22 MOhm serial resistor was added to limit eventual discharge currents. The light-source intensity was tuned with a series of absorbers placed in front of the lamp, adopting the light flux to the THGEM gain, within the above-mentioned current limits.

Except for the ETE, all measurements were carried out by recording the current from various electrodes in the different experimental setups. In most cases the current was measured on electrodes grounded through the precision electrometer (KIETHLY 610C), recording currents down to 10 pA; in some cases the currents on powered electrodes were measured indirectly by recording the voltage drop across a known resistor, which permitted measuring currents in the range of 10pA to 100 nA. The precision of these measurements was of 1% and 5%, respectively.

At very low THGEM voltages, below the multiplication threshold, the ETE can also be derived from the current measurements, by comparing $I_{OUT}$, the output current of the THGEM (i.e. collected on the interconnected THGEM bottom and mesh M2 electrodes – figure 3) to $I_{PC}$, the photocurrent emitted from the PC (measured at the PC with a field $E_{drift}$ established and no multiplication in the THGEM). But, as soon as the multiplication in the holes starts, this current measurement is no more valid for the ETE assessment; we cannot separate the ETE from the effects and charges resulting from the multiplication process [25]. In this range, the ETE was measured in a pulse-counting mode that permits separating the two processes. It is based on recording single-electron pulses, in which case electron transfer inefficiency is directly translated to counting-rate deficiency. We used a relative measurement, comparing the counting rate in the examined system to that recorded in a reference system known to have 100% ETE. This is done, of course, under exactly the same experimental conditions, with identical PC, UV-light illumination, and total pulse-gain and electronics chain. The pulse-counting method was used to obtain the transfer efficiency of the THGEM with either semitransparent or reflective PCs, in various gases. The technique details are given in [30,25] and in chapter 3 below.

## 3. Results and discussion

*3.1. MAXWELL & GARFIELD simulations*

MAXWELL and GARFIELD simulations were found to be very useful for understanding the role of the various geometrical parameters of the THGEM electrode and for comprehending their operation mechanism and the expected performance. A few examples will be given below.

MAXWELL calculation results of the electric field strength $E_{hole}$ along the hole's central axis, for THGEM#9 (Table I) with $\Delta V_{THGEM}$=2kV, is shown in fig.5. It reaches a maximum of ~40 kV/cm at the middle of the hole and remains above the multiplication threshold (10-15 kV/cm) along an additional ~ 0.3 mm distance outside the hole; it indicates that the gas multiplication will typically extend out of the hole under the maximal 2 kV bias. Other calculations showed that the avalanche will be fully confined within the hole at $\Delta V_{THGEM}$ =1.3kV [31]. A similar effect was noticed with a standard GEM in noble gases [32], showing evidence for the avalanche extending-out by much more than the hole radius.

Figure 6. Shows the results of MAXWELL calculation of $E_{hole}$ in a THGEM with t=0.4 mm, for different hole diameters. With decreasing hole diameter, $E_{hole}$ increases and becomes more confined within the hole. The resulting performance in terms of maximal $E_{hole}$ (and therefore the expected gain) shows an optimum for t/d= ~ 1, as will be discussed in the paragraph describing the gain results.

MAXWELL/GARFIELD calculations gave us another insight into the operation mechanism, as for example to the effect of the transfer field. In Fig. 4 the avalanche is simulated in a cascaded double-THGEM#9 in Ar/$CO_2$ (70:30), with $\Delta V_{THGEM}$=1350V on each multiplier and a high (3 kV/cm) transfer field between them. At $\Delta V_{THGEM}$=1350V the multiplication is~30 (this low gain was chosen for the sake of clarity of the figure); the total calculated gain is~900. This is a surprisingly high total gain, equal to the product of the two individual gains. With higher $\Delta V_{THGEM}$ and $E_{trans}$ values the calculated total double-THGEM gain exceeds the product of the two individual ones.

MAXWELL/GARFIELD provided the clue for this effect, showing that the high transfer field modifies the field near the hole's edge (fig. 7), thus modifying the multiplication factor. The effect is expected to be significant at the higher $\Delta V_{THGEM}$ values, where the avalanche further extends out of the hole. Furthermore, from GARFIELD calculations it is clear that a high transfer field is responsible for the efficient extraction of electrons from the first THGEM towards the second one in a cascade; the large hole size together with the extension of the field out of the hole is responsible for an efficient focusing of the electrons into the second THGEM. As will be shown in the next section this was confirmed experimentally.

The electric field on the top surface of the THGEM is shown in fig 8, along the line interconnecting two adjacent hole centers, for various $\Delta V_{THGEM}$ values. For $\Delta V_{THGEM}$ >800 V the field exceeds 3 kV/cm all over the surface. Under this relatively high electric field, in a multiplier layout with a *ref* PC, the photoelectron backscattering in the gas is low [29]; this guarantees its efficient extraction from the *Ref* PC into the gas.

*3.2 Effective Gain*

The experimental method for assessing the THGEM gain was explained above; the experimental schemes are given in [1] and in [24] for detector configurations with *Ref-* and *ST-* photocathodes and for double-THGEMs. The gain results are shown in figures 9-12 for various THGEM parameters and operating gases.

Fig. 9 demonstrates that a single THGEM provides up to a 10-fold higher effective gain than a standard GEM. The maximum gain, defined by the onset of sparks, is naturally reached at different $\Delta V_{THGEM}$ values, according to the multiplier's geometry. Fig. 10 shows the absolute effective gain of THGEM#9 in various gases; the highest effective-gain values, $\sim 10^5$, were reached in standard mixtures employed in GEMs; $CF_4$, which is an important gas for applications in windowless Cherenkov detectors

[29,33], yields a maximum gain of $10^4$, though at very high $\Delta V_{THGEM}$ values. Fig. 11 shows the gain of a double-THGEM#9 in Ar/CH$_4$ (95:5) at $E_{trans}$ =3kV/cm and in Ar/CO$_2$ (70:30), at $E_{trans}$ values of 1 and 3kV/cm; at 3kV/cm, the double-element multiplier yields up to 100-fold higher gains compared to that of a single-multiplier, reaching total effective gains of ~ $10^7$. Other electrodes were tested, e.g. THGEM#10, providing similar results. As will be discussed below in section 3.3, at the effective-gain values above a few hundreds, the ETE reaches 100% and therefore the effective gain is equal to the true gas multiplication factor within the holes.

The effect of $E_{trans}$ on the total gain, discussed in 3.1, is demonstrated in figure 11, showing that double-THGEM gains exceeding the product of two individual gains are obtained with a high $E_{trans}$ and high $\Delta V_{THGEM}$ values. This was tested in various gases and with different electrode configurations, showing systematically a similar behavior [31]. The double-THGEM structure provides very high total gains, while the voltages on each element are far from the sparking limit, which permits a spark-free operation. It was noted that the most spark-free double-THGEM operation is the symmetric one, namely with both elements biased at equal operation voltages. We also noted that the 0.1 mm etched Cu around the drilled holes is essential for achieving spark-free high gain. An attempt to operate a THGEM electrode, in which such etching was not done, resulted in ~10 times smaller maximal gain [24]. It is also important to have the etched and drilled patterns precisely centered. In our case the precision was ~ 20μm. Electrodes in which the etched pattern was largely displaced from the drilled one did not function properly, and yielded up to10-fold smaller maximal gain (fig 12).

The hole pitch of the THGEM was found to have a minimal effect on the gain. For example, the onset of the multiplication in THGEM#10, having a pitch of a=1mm, started a few tens of volts earlier than in THGEM#9, with a pitch a=0.7mm. Both multipliers reached similar maximum gains in Ar/CO$_2$ (70:30) (Figure 13). The effect was systematically observed also at the low-pressure range [24]. There is no clear explanation at the moment and it does not seem to be supported by our MAXWELL calculations.

The high gain obtained with the double-THGEM, permitted recording single-photoelectron signals with a fast current amplifier (Fig 14). The relatively fast multiplication process yields pulses with a few ns rise-time.

*3.3. Electron Transfer Efficiency*

We have measured the ETE and its dependence on the THGEM voltage, in two basic configurations:

a) *Ref PC,* in which the single electrons originate from a PC deposited on the top surface of the multiplier and the field $E_{drift}$ above it is set to 0. In these $E_{drift}$ conditions, like in a reflective-GEM [34,25], the photon detector has a considerably reduced sensitivity to ionizing particle background [33]. We also varied $E_{drift}$ around 0 and measured the resulting ETE variation.

b) *ST* PC, in which the single electrons originate from a PC placed a few mm above the THGEM electrode, with a field $E_{drift}$ between them; in this configuration the measured ETE is relevant for *ST* photon detectors (with $E_{drift}$ >0.5 kV/cm) and for tracking detectors and TPCs (with $E_{drift}$ typically in the range of 0.1kV/cm). It is also relevant for the understanding of the operation mechanism of two THGEMs in cascade, where avalanche electrons created in the first multiplier should be efficiently focused into the second one.

Fig. 15 depicts the setup and method for ETE measurement with a *ref* PC, which has two steps: first we set $E_{drift}$ = -3kV/cm and measure the event rate originating from electrons created at the *Ref* PC and multiplied at the $MW_{nor}$ anode. The high $E_{drift}$ ensures full photoelectron extraction efficiency; the electric field established on the MW side of the mesh M1 is higher than 6kV/cm, ensuring full electron transfer through M1. We may thus assume that in this configuration the ETE is 1. Then, with the same light flux and electronics chain, we set $E_{drift}$= 0 and measure the event rate originating from electrons entering the THGEM and being multiplied in a cascade: first in the holes and further on $MW_{trans}$ anode. This two-stage multiplication arrangement permits varying the THGEM gain while keeping a fixed total gain on the cascade. The ratio of event rates ($n_{trans}/n_{nor}$) provided us with the ETE.

The validity of the measurement relies on the assumption that in both cases the single-electron pulse-height distribution is exponential, following the Polya relation without saturation

$$P(q) \cong (q/Gq)e^{-(q/G)} \qquad (1)$$

G being the gain, q being the amplitude [35]. Therefore it is important to adjust the total gain in both measurement steps to be identical within 2-5%, by comparing the slopes of the exponential distributions. Furthermore, we measured the event-rate within a given window (Figure 16), set in the middle of the pulse-height distribution, safely above the noise and below the tail, to avoid counting secondary or pile-up pulses. The method is no more valid in cases where the multiplication process is strongly affected by secondary or quenching processes and the distribution fails to follow the exponential relation. For a more detailed discussion of this method refer to [25,30,34].

The results of the ETE with *Ref* PC on a THGEM#9 in four gases investigated in this work, are shown in fig.17 as function of $\Delta V_{THGEM}$. With $CF_4$, in which multiplication starts at very high voltages (see fig. 10), ETE was evaluated by current measurement up to ~1400V and by pulse counting in the range above that. Full transfer efficiency is obtained at rather low gains, of 3-30, according to the gas filling. (Note that for ETE=1 $G_{eff}$ equals true gas multiplication). This could be compared to a standard *reflective* GEM, in which full ETE was attained only at high gains, above 500 in $Ar/CH_4$(95:5) and above 5000 in pure $CF_4$ [25]. The reason is the denser hole area (46% of the area, compared to 22% in a standard GEM) and the larger hole diameter (300 μm compared to 50-70 μm in a standard GEM). Due to the large hole diameter, which is indeed larger than the electron diffusion (~100μm for 1 cm [36]), electron focusing into the holes is more efficient and is typically obtained at smaller field (i.e. gains) compared to that of a standard GEM. The ETE of a better-suited multiplier for *Ref* GPMs, THGEM#10, which has a higher effective PC area of 77% (similar to that of a standard GEM), is shown in Figure 18. Due to the larger hole distance in this case, a higher gain of ~500 is required for full electron transfer efficiency.

In fig. 19 we show the dependence on $E_{drift}$ of ETE of THGEM#9 with *ref PC* in $Ar/CH_4$ (95:5). Like in a GEM [34,25] full transfer efficiency was measured for

$E_{drift}$=0. Setting $E_{drift}$ at slightly reversed (negative) value will reduce the detector's sensitivity to ionizing background, as all ionizing electrons will drift away from the multiplier.

Figure 20 depicts the experimental setup and method for measuring ETE with *a ST* PC. The setup included two 20nm thick CsI layers deposited on both faces of a thin quartz plate, pre-coated with 2.5nm thick Cr. Similarly to the *Ref* PC mode described above, we had a normalization and a measurement step, and we used the ratio of event rates in both steps to provide the ETE. However, the normalization was done in two steps. First we recorded, under the same UV illumination and the same extraction field $E_{drift}$, the photocurrents from both sides of the quartz plate; this provided us with the photocurrents ratio $R_I$ between the top *Ref* PC and the bottom *ST* PC. Then we preceded as above and measured the rate of events recorded in the defined pulse-height window, for events originating from the top *Ref* PC and amplified in the top $MW_{nor}$. Finally the event rate was measured within the same pulse-height window and under the same illumination, for events originating from the *ST* PC and multiplied in the THGEM and the $MW_{trans}$ in cascade, maintaining the same total detector gain and electronics chain. The ratio of the two event-rates, normalized by the photocurrents ratio $R_I$, provided us with the ETE, as function of $\Delta V_{THGEM}$ and of $E_{drift}$.

The ETE results in the *ST* PC mode for THGEM#9 are shown in figure 21 as function of $\Delta V_{THGEM}$, for two gases. These data were obtained by the current-recording method in the voltage range below the multiplication onset, and by the pulse-counting method in the multiplication range. Full transfer efficiency was attained in Ar/$CO_2$ (70:30) and in pure $CH_4$ already at small respective gains of 100 and 10, with $E_{drift}$= 0.3 V/cm. (Note that for ETE=1 $G_{eff}$ equals true gas multiplication). As in standard GEM the electron focusing into the holes, and thus the ETE, is expected to drop when the ratio $E_{drift}/E_{hole}$ increases. The ability to maintain full ETE at higher $E_{drift}$ was measured for THGEM effective gains of 10, $10^3$ and $10^4$, as seen in figure 22. A drop is observed at $E_{drift}$ values above ~3kV/cm and ~5kV/cm for effective gains of 10 and $10^3$-$10^4$, respectively.

As discussed above, the ETE measured with a *ST* PC is relevant also for the operation of a cascaded-THGEM structure. The results of fig. 22 confirm that even with transfer fields between two cascaded elements as high as 3kV/cm, a full electron focusing into the second THGEM holes can be obtained.

In analogy to standard GEM operation in cascade, the charge transferred to the second element depends not only on the ETE discussed above but also on the electron extraction efficiency from the first multiplier into the gap between them. This efficiency is expected to increase with $E_{trans}/E_{hole}$. Its dependence on $E_{trans}$ was measured in a double THGEM configuration similar to that shown in fig. 4, with a ST PC. First we measured the current $I_B$ collected at the bottom of THGEM1, with a reversed $E_{trans}$, and then we measured the current $I_T$ on the top of THGEM2, with its top and bottom interconnected, as function of $E_{trans}$. The ratio $I_T/I_B$, provides the electron extraction efficiency from THGEM1 to the gap between the THGEMs. Experimental data are shown in figure 23 for $Ar/CO_2$ (70:30). At effective gain of $10^4$ full extraction efficiency from THGEM1 in this gas is achieved at $E_{trans}>6kV/cm$ and 65% at $E_{trans}$ of 3 kV/cm.

*3.4 Counting rate capability*

The pulse-height dependence on the event rate is important for high-rate applications. Due to the reduced number of holes per $mm^2$ compared to standard GEM, each hole contains higher electron (and ion) flux, which could be of a concern.

The measurements were done in two steps, in a setup similar to that of figure 3, with a *Ref* PC deposited on a THGEM#10, and a mesh M1 placed a few mm above it. A collimated UV lamp illuminated a PC area of 7 $mm^2$. The current limit on the power supplies was raised to 500nA and all current-limiting resistors were removed. First, the photocurrent $I_0$ was measured on M1 as a function of the UV intensity with both sides of the THGEM interconnected and with $E_{drift}$=3kV/cm. This provided the photoelectron rate per unit area. Then $E_{drift}$ was set to 0, the THGEM was biased with $\Delta V_{THGEM}$ to a known gain and the current $I_1$ was recorded at the THGEM bottom, with a reversed $E_{trans}$, again as function of the UV intensity. $I_1/I_0$ provided the gain of THGEM#10 and its dependence on the impinging photoelectron flux.

The results are shown in Fig. 24 for two different gains, in Ar/$CO_2$ (70:30). At gain of $2\times10^4$ (maximal gain in this gas) a multiplication drop starts at ~$10^7$ electrons/mm²sec.

The results could be compared to that of a standard GEM operated at a gain of $10^4$, irradiated with 5.9keV x-rays, where the pulse-height was constant up to a total event rate of $10^5$ converted x-rays/mm² sec [37]. Assuming about 250 electrons per x-ray this corresponds to ~ $2.5*10^7$ electrons//mm² sec. This very high rate capability, e.g. a few orders of magnitude higher than in wire chambers, could be of prime importance in some applications.

*3.5 Ion Backflow Fraction (IBF)*

IBF is relevant both for TPCs, where the ions are causing dynamic field distortions and for gaseous photomultipliers incorporating a solid PC, where the ions create PC physical and chemical aging and induce secondary electron emission causing feedback pulses that limit the detector performance. A comprehensive discussion on the ion backflow in gaseous detectors, its consequences and methods for its reduction is given in [20].

In the present work we have measured IBF for single- and double-THGEM structures, with *ST* PCs; the data is relevant for TPCs.

The IBF for a single THGEM#9 and a *ST* PC was measured in a setup similar to figure 3, with the THGEM bottom and M2 electrodes interconnected and with varying $E_{drift}$. The IBF was deduced from the ratio of currents recorded on the PC and on the interconnected THGEM bottom and M2 electrodes. The results are shown in figure 25 as function of the field $E_{drift}$ above the THGEM. The fraction of avalanche-induced ions which drift towards the photocathode is less than 2% at $E_{drift}$=0 and increases almost linearly with $E_{drift}$. These data imply that in double-THGEM operation with $E_{trans}$~3kV/cm between both THGEMs, less than 40% of the ions will be flowing from the second THGEM towards the first one. Out of this, part may be trapped at the first THGEM bottom electrode, thus reducing the IBF as compared to that in a single THGEM operation. The IBF graph shown in Fig. 25 was measured at a gain of $10^4$.

The IBF with double THGEM#10 and a semitransparent PC was measured in a setup similar to figure 4, in 760 Torr Ar/CO2 (70:30) for two values of $E_{trans}$ and with $E_{drift}$=1.2kV/cm. Fig. 26. The IBF drops with increasing THGEMs voltage. This is due to the increasing lateral spread of the avalanche, resulting in the majority of ions being created at points far from the hole axis. With a sufficiently large hole dipole-field these ions are diverted and trapped on the top of THGEM2 and on the bottom of THGEM1.

*3.6. Energy resolution*

The energy resolution of the THGEM#9 was assessed with 5.9 keV $^{55}$Fe x-rays in 740 Torr Ar/CH$_4$ (95:5); the source irradiated an area of ~7 mm$^2$. A conversion gas gap of 8.5 mm was added in front of the multiplier, with a drift field set to 1.25kV/cm. The detector was operated at a gain of $10^5$. Pulses from the bottom electrode of the THGEM were recorded, via a charge-sensitive preamplifier (ORTEC 142) and a linear amplifier (ORTEC 570), on a multi-channel analyzer (fig 27); the resolution is ~20% FWHM. For comparison, a resolution of 18% FWHM was recorded with 5.9 keV x-rays in a standard GEM at a gain of 1000 in Ar/DME(80:20) [38].

**4. Summary**

The THGEM discussed in this article is an attractive robust and economic electron multiplier, suited for applications at atmospheric gas pressure requiring large area detectors with single-electron sensitivity and moderate (sub-mm) localization resolution. Possible applications could be in large TPC readout or in sampling elements in Calorimetry. The high attainable gains, of $10^4$ –$10^5$ in a single multiplier and 10-100 times higher in a double-THGEM multiplier, are due to the large hole size, reduced photon-feedback and efficient electron transport processes. The THGEM has spark-free operation in a variety of gases including pure CH$_4$ and CF$_4$. The rapid avalanche process developing across the hole results in fast signals, of few ns rise-time; the counting-rate capability is up to the range of ~10MHz/mm$^2$ at a gain of 10$^4$. THGEM multipliers can be coupled to both gaseous ionization volumes and to solid radiation converters; in the latter configuration the converter can be placed above the THGEM or deposited directly on its top face. In both cases the radiation-induced emitted electrons are efficiently focused into the multiplication holes. X-rays were detected with a gas converter with an energy resolution of ~20% FWHM at 5.9 keV. The solid converter material can be chosen according to the application; it can be a photocathode in gaseous photomultipliers [16], an x-ray converter (e.g. CsI) in secondary-emission x-ray imaging detectors [39] or a neutron converter (e.g. Li, B, Gd, polyethylene etc.) in thermal-or fast-neutron imaging detectors [15]. Such detectors for fast-neutron imaging are under development at our group.

From our systematic study we may conclude that the operation mechanism, as well as the role of the various electric fields involved in the THGEM operation, is rather similar to that known for standard GEMs.

In particular, we observed the following similarities and differences:

- The maximal voltage difference across the THGEM before sparks onset does not scale with the dimensions and the field inside the holes is smaller than in a GEM; but due to the larger dimensions, particularly the larger thickness, significantly higher gains are obtained. Furthermore, due to the larger hole-size (larger than the electron diffusion) electron focusing into the holes is more efficient and is typically obtained at smaller gains compared to that of a standard GEM.

- In our study of the role of each field we have confirmed that with a *Ref* PC the field $E_{drift}$ above the THGEM should be 0, to reach maximum electron focusing into the holes; it can be kept slightly reversed to reduce the sensitivity of detectors with solid converters to ionizing background.

- Unlike a GEM coupled to a *ST* PC, in which $E_{drift}$ should be kept moderate to avoid diverting the drifting electrons towards the metallic GEM surface, the large holes in the THGEM permit an operation with very high $E_{drift}$ values; an efficient photoelectron focusing into the holes even at drift fields of 5kV/cm was measured with THGEM#9 at a gain of 10$^4$. This is important for the efficient extraction of

- photoelectrons or radiation-induced secondary electrons, particularly in noble-gas mixtures, where backscattering into the converter is high at low fields [40].
- The dipole field within the holes deflects the avalanche electrons towards the bottom face of the THGEM, but with a strong $E_{trans}$ underneath the THGEM electrode the charge is efficiently diverted and transferred into the following multiplier in the cascade. In standard cascaded GEMs this is an important issue of optimization, since with a too high $E_{trans}$ the electrons will not be focused into the second GEM, while a too low $E_{trans}$ will not extract the electrons from the first one [25]. With the large holes, electron focusing into the second THGEM in a cascade remains effective even with very high $E_{trans}$ values, as with the high $E_{drift}$ values discussed above.
- As a result, a double-THGEM operation was proved very efficient and spark-free in all tested gases, providing high total gains. Very high $E_{trans}$ values, of several kV/cm, could be applied to increase the transfer efficiency and thus the total gain. In some cases, at high THGEM voltages and with a high $E_{trans}$ value between the two elements, the total effective gain exceeded the product of the two individual ones. This peculiar feature is occurring firstly because the extraction of charges from the holes into the next stage is very efficient and reaches almost 100%; furthermore, because the dipole hole-field is extending out by about the hole's diameter, at large gains part of the avalanche is developed outside the hole, thus being susceptible to modifications by any strong field in the gaps around the THGEM. The extension of the avalanche outside the hole might have undesirable consequences, such as instability due to photon-mediated processes. Thus care should be taken to choose the appropriate operation conditions, such as THGEM geometry and applied voltages, specific for each application, in order to avoid this phenomenon.
- The electron transfer efficiency (ETE) with a reflective PC deposited directly on the top of the THGEM#9 was found to reach ~100% at rather low THGEM voltages (i.e. THGEM gain), in all gases studied. However, with THGEM#10, of larger pitch and of hole area similar to that of a standard GEM, higher THGEM voltages (i.e. THGEM gains) were required to attain full ETE. This is in fact expected, because the ETE in this case includes also the electron extraction efficiency from the PC to the gas, which requires high field (>0.5kV/cm) on the PC surface. With the larger pitch higher THGEM voltages are therefore required.

- Just as in a standard GEM, the flow of back-drifting ions is strongly related to that of the avalanche electrons. THGEMs seem to have efficient electron transport, and not surprisingly also efficient ion transport, as compared to GEMs. However, with THGEM#10 (optical transparency identical to standard GEM), IBF is typically slightly below 10% with fields suitable for *ST* PC operation. This result is quite similar to that obtained with 3GEM cascade [42] under similar conditions. Reducing ion backflow in THGEM cascades is the main subject of our ongoing research, attempting to apply the idea of reversed-bias multi-hole & strip electrodes (R-MHSP), recently shown [20] to reduce the ion backflow fraction by a factor of ~$10^3$ when incorporated in a cascaded structure.

In summary, the THGEM can be easily produced, spanning a large scale of geometrical parameters: we have tested such electrodes with thickness ranging from 0.4mm up to 3.2mm, and with hole sizes and distances in the same range. By varying the thickness and the hole size it is possible to optimize the THGEM for various operation conditions, as for example the operation at very low gas pressures, in the mbar range [1,24]. Similarly, varying the holes pitch affects the ETE and thus permits optimization of the electrode to a particular operation layout. The holes pitch also affects the localization precision provided by this electrode, and some optimization with regards to the localization demands are also possible here. The gain homogeneity and the localization properties of 100x100 mm$^2$ THGEM detectors are the subject of another work.

Though the gain in single-and double-THGEMs is high, further improvements could be achieved by using different geometries. One example is using smaller hole diameter, where according to calculations, $E_{hole}$ can reach much higher values. Conical holes shape, like in the standard GEM, may also be tested. For UV-photon imaging in RICH applications, where the detector operates in $CF_4$, the G-10 substrate failed reaching high gains following discharges [23]. This was attributed to possible damages caused by Fluor radicals to the glass fibers of the G-10 material. An attempt to use copper clad Kevlar instead of G-10 did not provide satisfying results so far. Other materials, less sensitive to $CF_4$, should be tested. The operation of THGEM imaging detectors under high radiation flux is the subject of an undergoing research. Further studies on the THGEM should include the investigation of temporal gain stability and its relation to up charging of the electrode due to the 0.1mm wide insulator within the high electric field. Some evidence of temporal instability have been observed [42], which calls for optimization of the insulator width.

**Acknowledgments**

We would like to thank Mr. D. Shafir for his contribution. This work was supported in part by the Benoziyo Center for High Energy Reserach, the Israel Science Foundation, project No151-01 and by the Binational Science Foundation project No 2002240. C. S. was supported in part by the Fund for Victims of Terror of the Jewish Agency for Israel. A.B. is the W.P. Reuther Professor of Research in the peaceful use of Atomic Energy.

*Table I. A summary of the geometrical parameters of THGEMs studied in this work.*

| THGEM# | Thickness t [mm] | Drilled hole diameter d [mm] | Etched Cu diameter [mm] | Pitch a [mm] | *Ref* PC area [%] | Low (L) or Atm (A) pressure |
|---|---|---|---|---|---|---|
| 1 | 1.6 | 1 | 1 (no etching) | 7 | 98 | L* |
| 2 | 1.6 | 1 | 1 (no etching) | 4 | 94 | L* |
| 3 | 1.6 | 1 | 1.2 | 4 | 92 | L* |
| 4, 6 | 1.6 | 1 | 1.2 | 1.5 | 42 | L*+A |
| 5 | 3.2 | 1 | 1.2 | 1.5 | 42 | L* |
| 7 | 0.4 | 0.5 | 0.7 | 1 | 56 | A |
| 8 | 0.8 | 0.5 | 0.7 | 1 | 56 | A |
| 9 | 0.4 | 0.3 | 0.5 | 0.7 | 54 | A |
| 10 | 0.4 | 0.3 | 0.5 | 1.0 | 77 | A |
| 11 | 2.2 | 1 | 1.2 | 1.5 | 42 | L* |
| Standard GEM | 0.05 | 0.055 | .07 | .14 | 77 | |

*Results presented in the second, related article on low-pressure THGEM operation [24].

**Figure Captions**

Figure 1. A microscope photograph of a THGEM with thickness t = 0.4 mm, hole diameter d= 0.3 mm and pitch a= 0.7 mm. A rim of 0.1 mm is etched around the mechanically drilled holes.

Figure 2. The THGEM operation principle demonstrated by GARFIELD simulation at low gain (~30): electrons from gas ionization or from a photocathode (semitransparent or reflective) are focused into the holes and multiplied by an avalanche process under $E_{hole}$. Depending on the direction of $E_{trans}$, avalanche electrons are transferred to a second multiplier/readout electrode or collected at the THGEM bottom electrode, as shown here with a reversed $E_{trans}$.

Figure 3. Electric field map in THGEM#9 (table I) calculated by MAXWELL for $\Delta V_{THGEM}$=2kV, $E_{drift}$=0.1kV/cm and $E_{trans}$=3kV/cm; the latter are set by the potentials on meshes M1 (or photocathode) and M2. The calculated field $E_{hole}$ within the holes varies between 10 - 40kV/cm.

Figure 4. GARFIELD simulation of the avalanche process in a double-THGEM#9, atmospheric pressure Ar/$CO_2$ (70:30), $\Delta V_{THGEM}$=1350 V. The multiplication factor of each THGEM is ~30, resulting in a total gain of ~900. With reversed $E_{trans}$ all electrons are collected at the bottom of the second THGEM.

Figure 5. $E_{hole}$, the electric filed strength along the central hole axis, calculated by MAXWELL in a standard GEM and in a THGEM#9, for maximal operation voltages in Ar/$CO_2$ (70:30): $\Delta V_{THGEM}$=2kV; $\Delta V_{standardGEM}$=0.5kV.

Figure 6. $E_{hole}$, the electric field strength along the central hole axis, calculated by MAXWELL for a fixed plate thickness t=0.4 mm and different hole diameter d. With decreasing hole diameter the field increases and is more confined within the hole. Experimental measurements show maximum gain for t/d ~ 1 (see figure 9 and text).

Figure 7. The effect of $E_{trans}$ on the electric field strength $E_{hole}$ along the hole axis, calculated by MAXWELL. As seen on the left side of the figure, a high $E_{trans}$ applied above or below the THGEM may significantly affect the field strength just outside the hole and therefore the multiplication process.

Figure 8. The electric field on THGEM#9 top surface, $E_{surface}$, calculated by MAXWELL along the line interconnecting two hole centers. The electric field magnitude is above 3kV/cm, even at $\Delta V_{THGEM} = 800V$.

Figure 9. Absolute effective gain measured with THGEMs of different geometries and with a standard GEM, in 740 Torr Ar/CH$_4$ (95:5). The highest gains were recorded with THGEMs having t/d= ~1.

Figure 10. Absolute effective gain of THGEM#9, measured in different gases at 740 Torr.

Figure 11. Absolute effective gains of single- and double-THGEM#9 multipliers, measured in Ar/CH$_4$ (95:5) with $E_{trans}$ =3kV/cm, and in Ar/CO$_2$ (70:30) with $E_{trans}$ of 1 and 3kV/cm.

Figure 12. A THGEM#7 electrode with etched holes displaced by ~40 - 50 μm relative to the drilled ones has significantly lower maximal gain.

Figure 13. Varying the hole pitch from 0.7 to 1 mm has a minor effect on the gain curve.

Figure 14. A fast single-photon pulse, of 8 ns rise-time, measured in double-THGEM#7 in 740 Torr Ar/CH$_4$ (95:5) at a gain >$10^6$.

Figure 15. The experimental setup for the Electron Transfer Efficiency measurements in a reflective-photocathode mode. The pulses were measured on the upper, $MW_{nor}$, and on the lower, $MW_{trans}$, multiwire detectors.

Fig. 16: Single-photoelectron spectra of the normalization and electron-transfer measuring steps, recorded on $MW_{nor}$ and $MW_{trans}$, respectively. The integration window is indicated.

Figure 17. ETE measured with THGEM#9, as function of $\Delta V_{THGEM}$ in a reflective photocathode mode, in various gases; full transfer efficiency is achieved at significantly lower gains compared to the standard GEM.

Figure 18. ETE measured with THGEM#9 and THGEM#10, as function of $\Delta V_{THGEM}$ in a reflective photocathode mode, in Ar/CO$_2$ (70:30).

Figure 19. ETE measured with THGEM#9 at $G_{eff}=10^3$, as function of $E_{drift}$ in Ar/CH$_4$(95:5); maximum ETE was obtained at $E_{drift}=0$.

Figure 20. The experimental setup for the Electron Transfer Efficiency measurements in a semitransparent photocathode mode.

Figure 21.   ETE of THGEM#9 in the semitransparent photocathode mode, in $Ar/CO_2$ (70:30) and in pure $CH_4$. Full ETE is reached at respective effective gains of $10^2$ and 10, with $E_{drift}$=0.3kV/cm.

Figure 22.   ETE of THGEM#9 in the semitransparent photocathode mode, as function of $E_{drift}$; $Ar/CO_2$ (70:30) at different effective gains.

Figure23.   Electron extraction efficiency from THGEM#9 to the gap below it, at effective gains of 10 and $10^4$ in $Ar/CO_2$ (70:30), as function of $E_{trans}$.

Figure 24.   Counting rate response of a single THGEM#10 with reflective photocathode in $Ar/CO_2$ (70:30). At effective gain of $2x10^4$ multiplication drop starts at ~$10^7$ electrons/mm$^2$sec.

Figure 25.   The Ion Backflow Fraction to a semitransparent photocathode, of a single THGEM#9 at effective gain of $10^4$, as function of the field $E_{drift}$ above the THGEM.

Fig 26.   The Ion backflow fraction to a semitransparent photocathode, of double THGEM#10, as function of the total effective gain, for two values of $E_{trans}$ and with $E_{drift}$=1.2 kV/cm. The IBF is below 10% in this case.

Figure 27.   An energy spectrum recorded with 5.9 keV $^{55}$Fe X-rays in a single-THGEM#9, 740 Torr $Ar/CH_4$ (95:5), effective gain $10^5$.

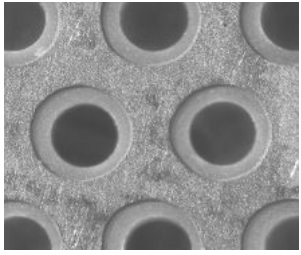

*Figure 1.*

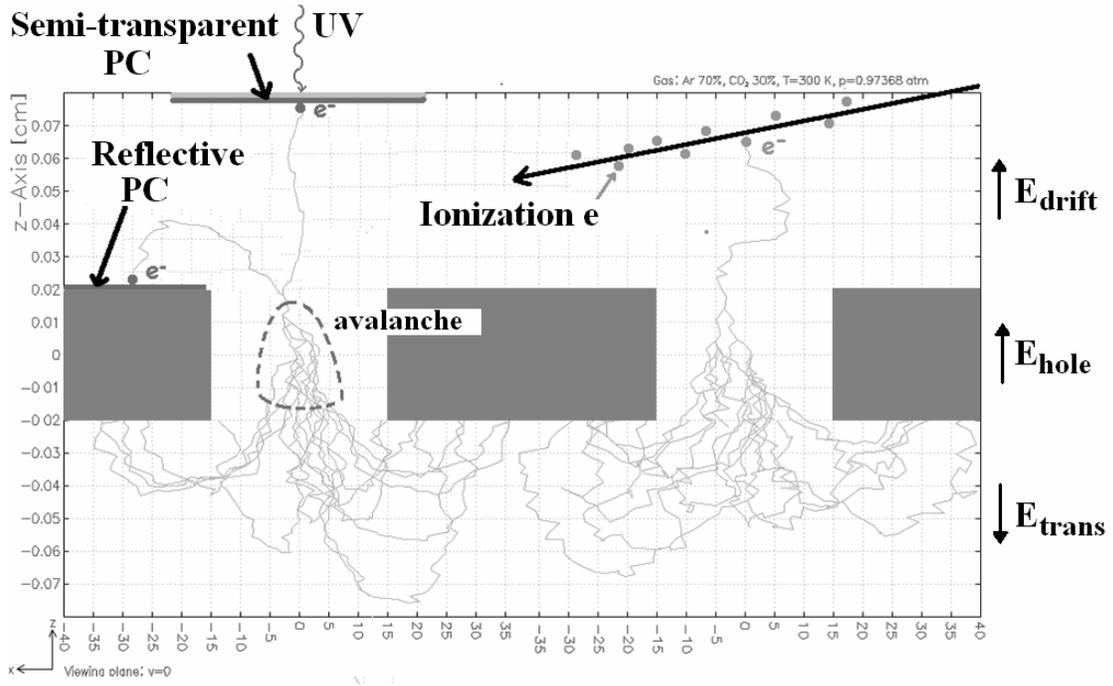

*Figure 2.*

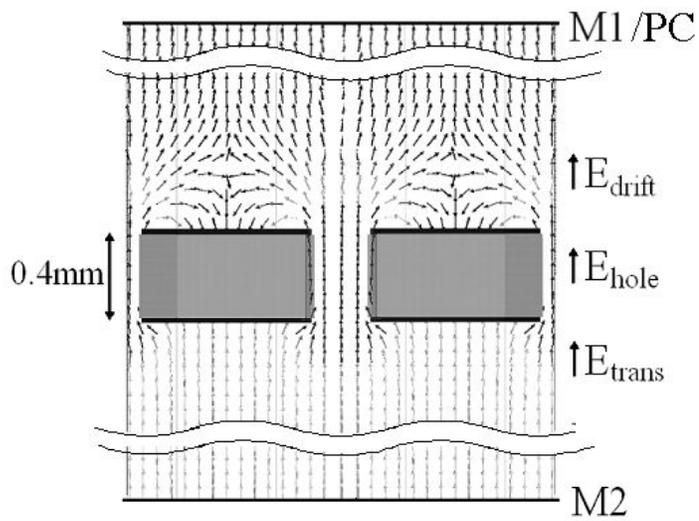

*Figure 3.*

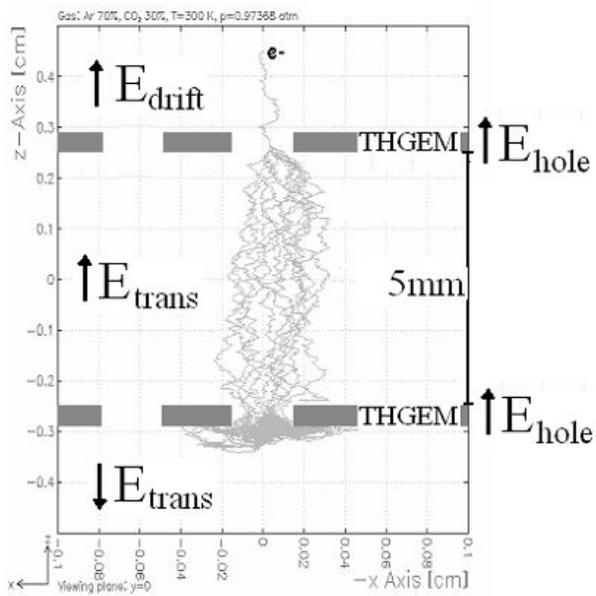

*Figure 4.*

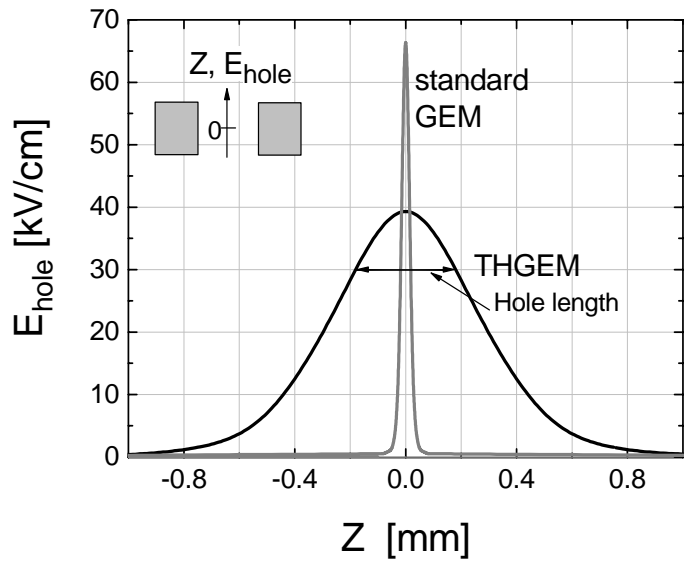

*Figure 5.*

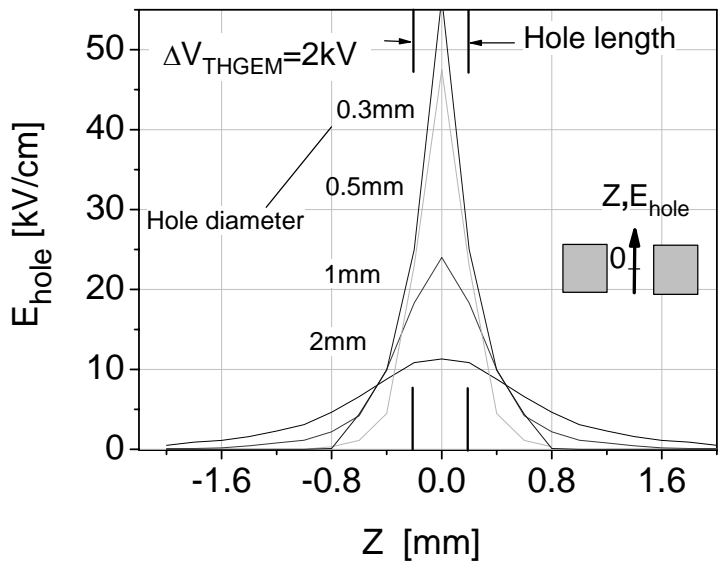

*Figure 6.*

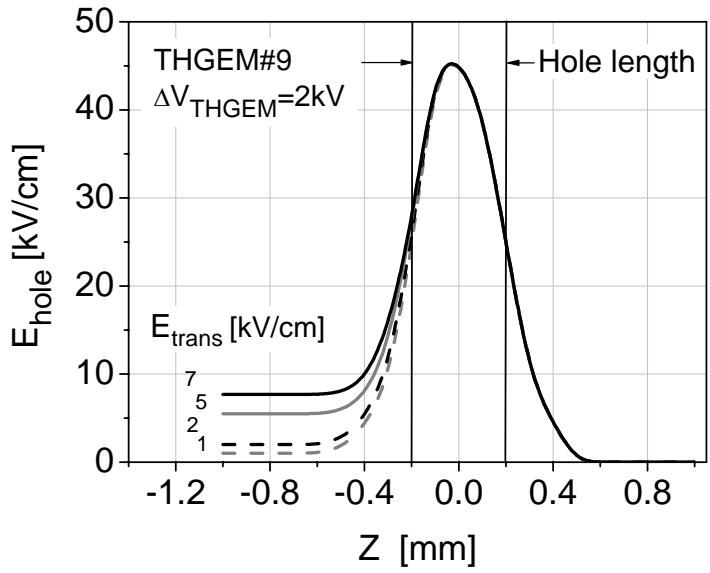

*Figure 7.*

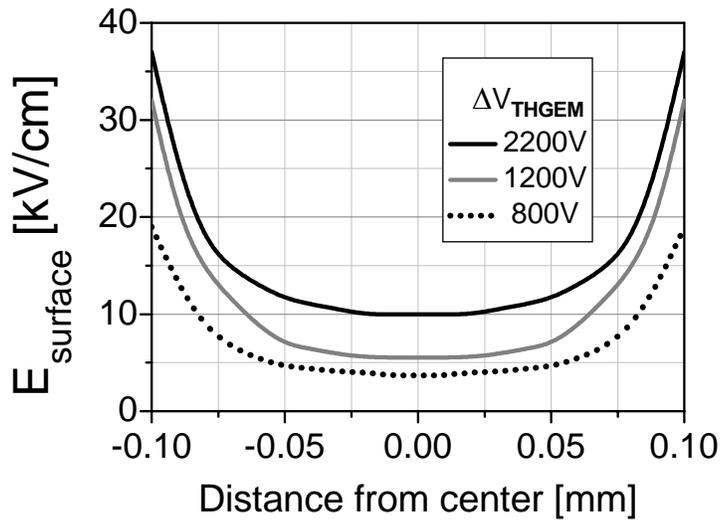

*Figure 8.*

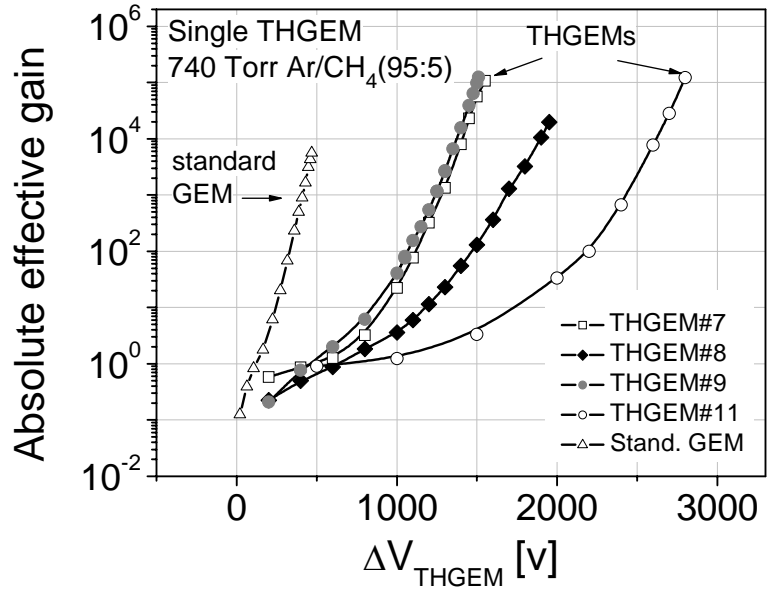

*Figure 9.*

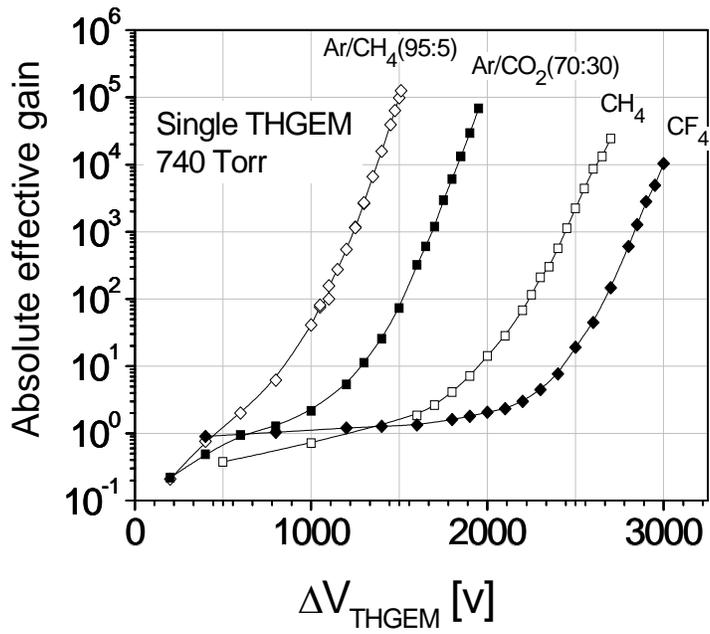

*Figure 10.*

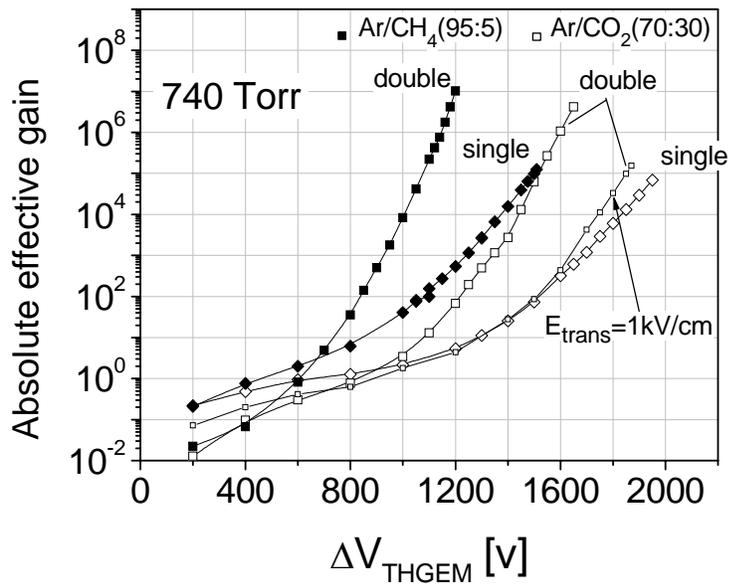

*Figure 11.*

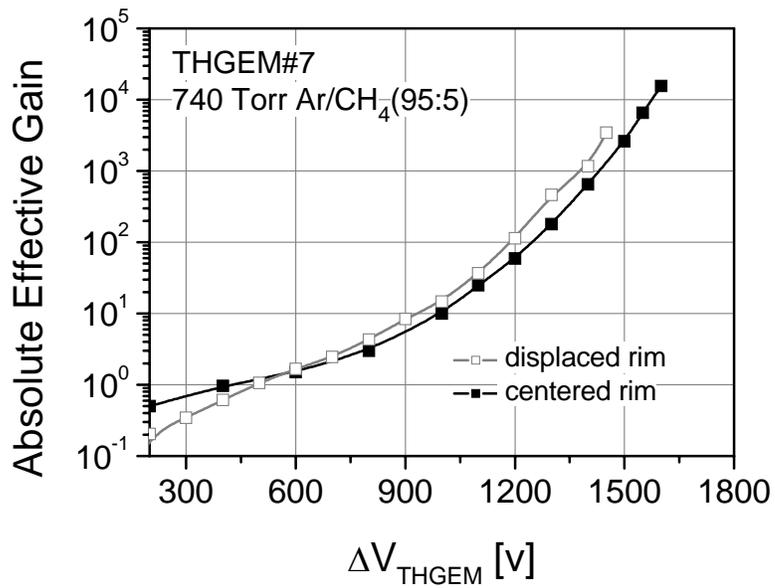

*Figure 12.*

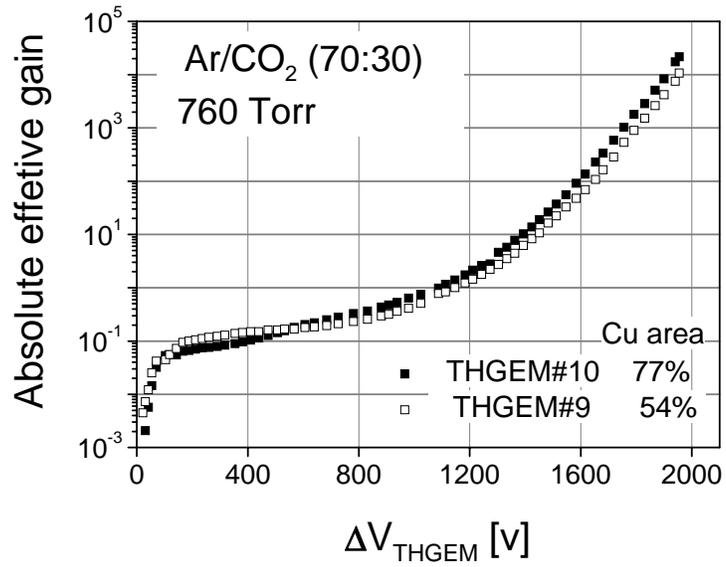

*Figure 13.*

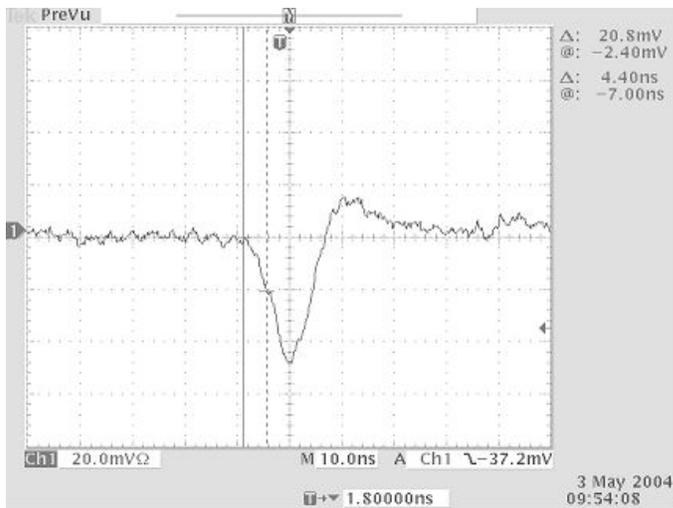

*Figure 14.*

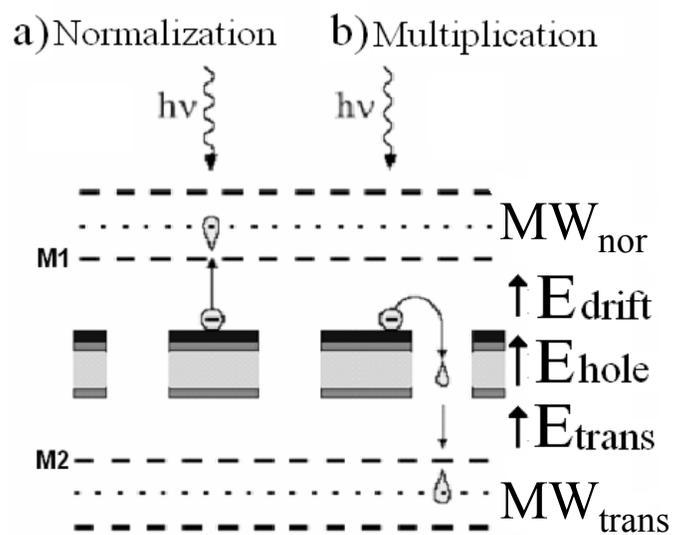

*Figure 15.*

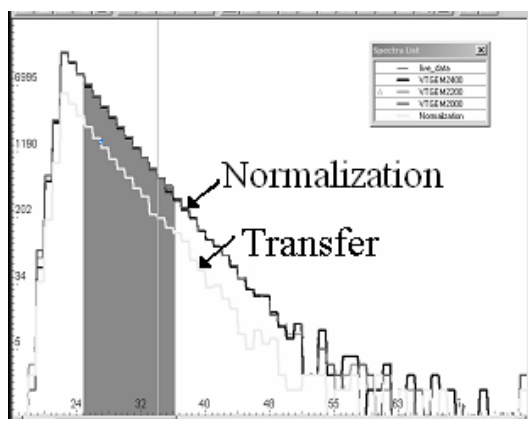

*Fig. 16:*

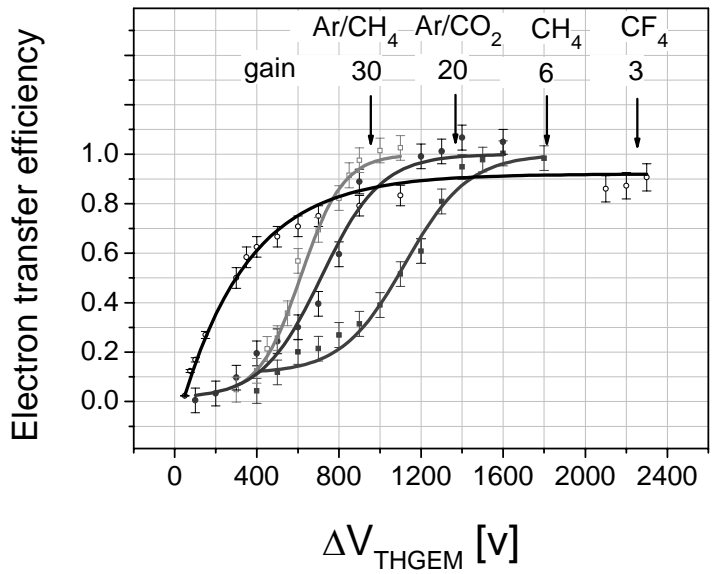

*Figure 17.*

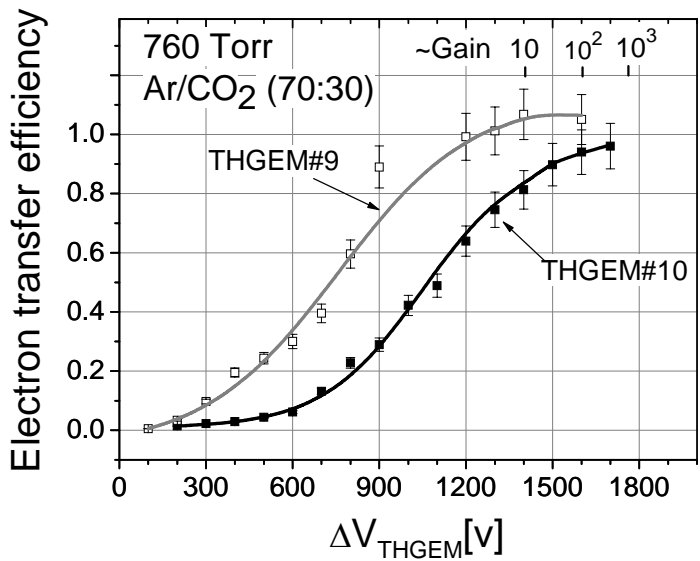

*Figure 18.*

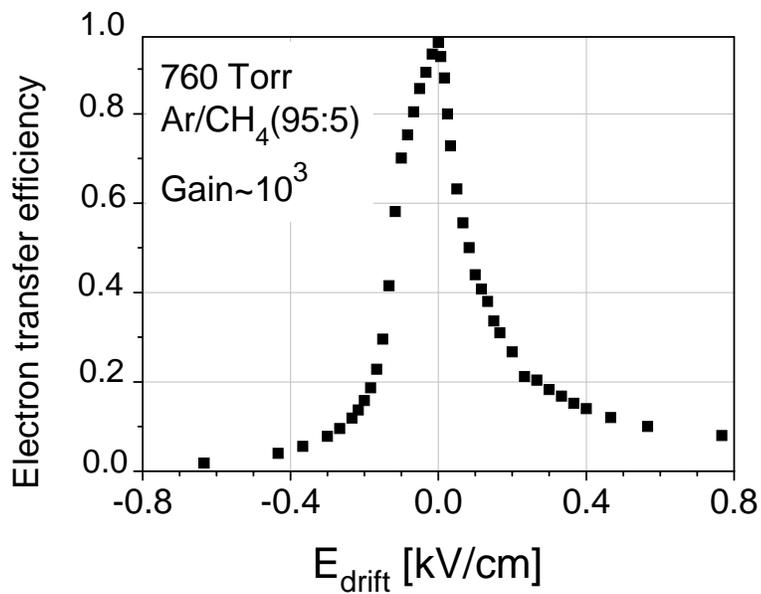

*Figure 19.*

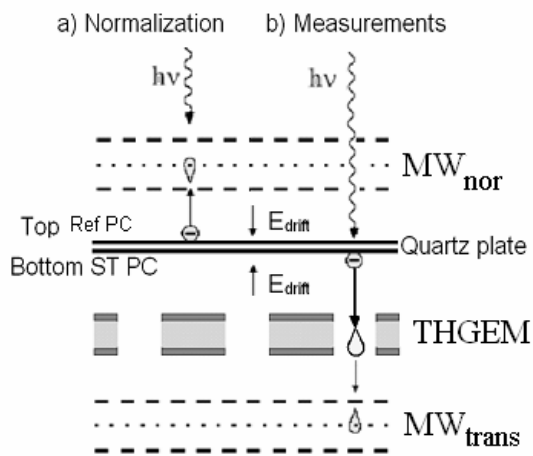

*Figure 20.*

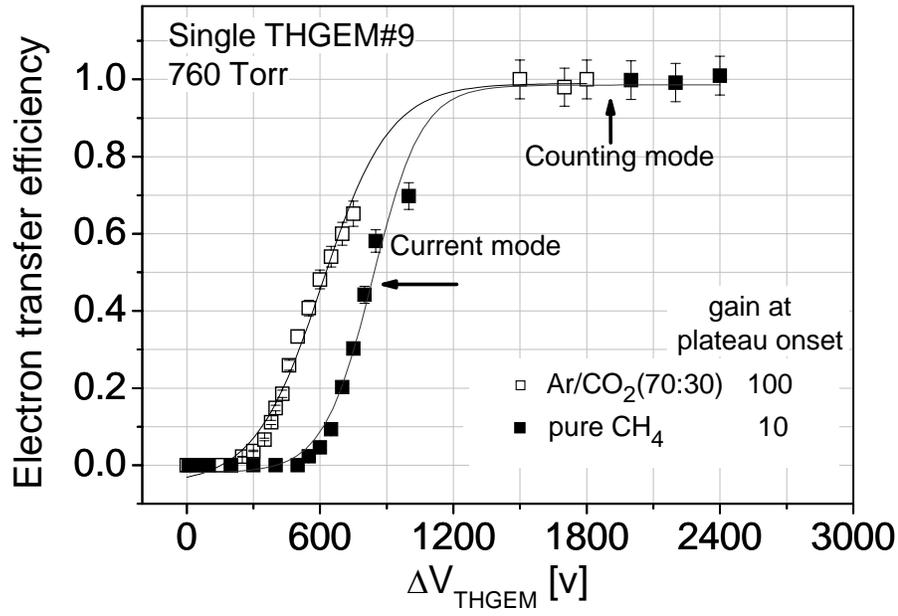

*Figure 21.*

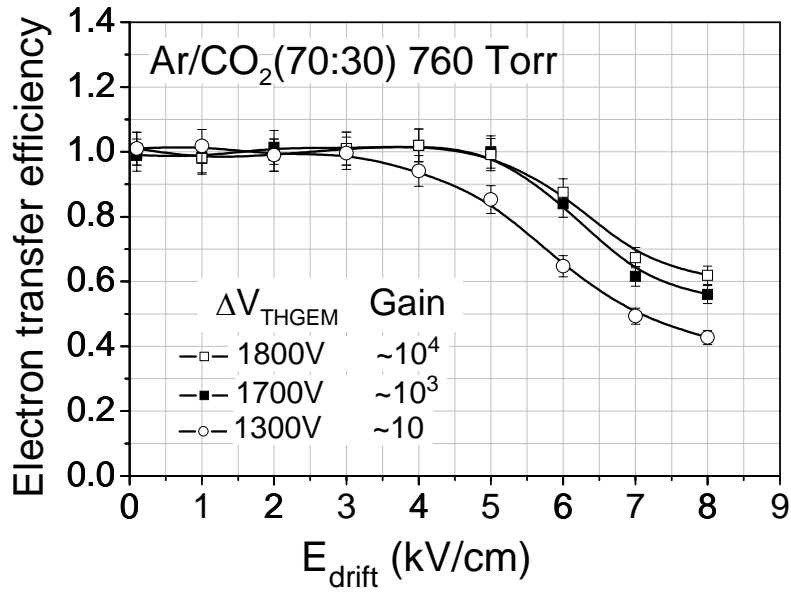

*Figure 22.*

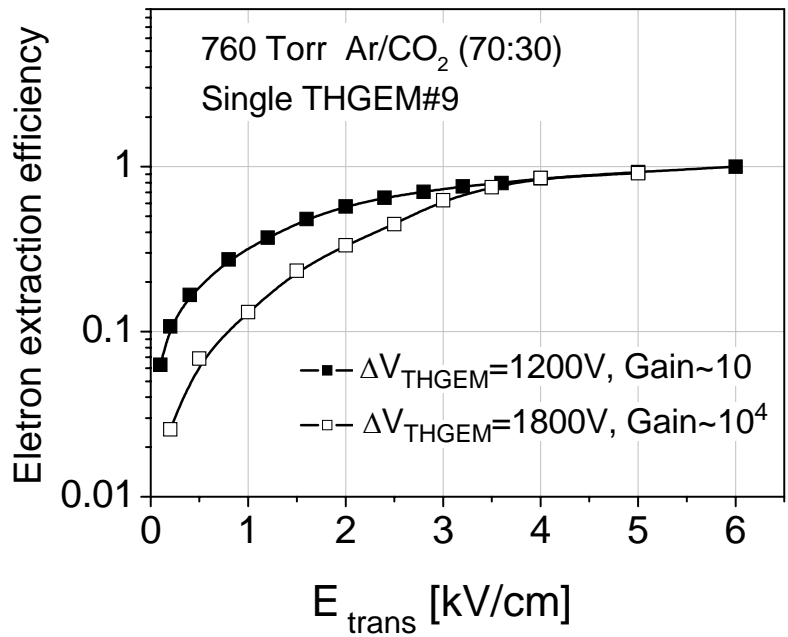

*Figure23.*

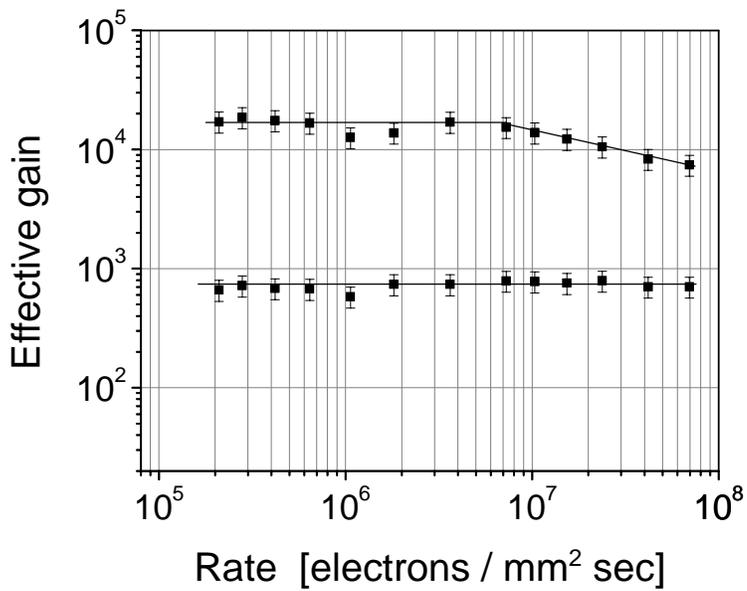

*Figure 24.*

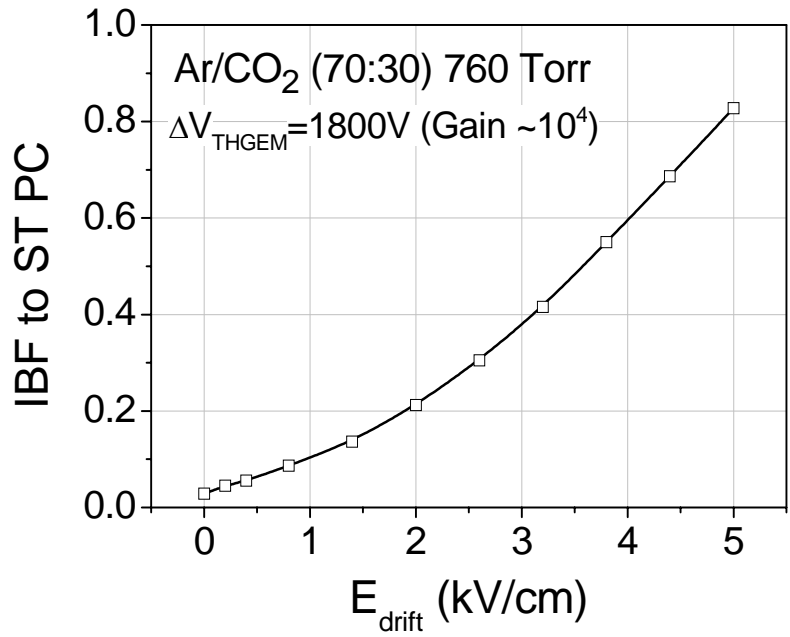

*Figure 25.*

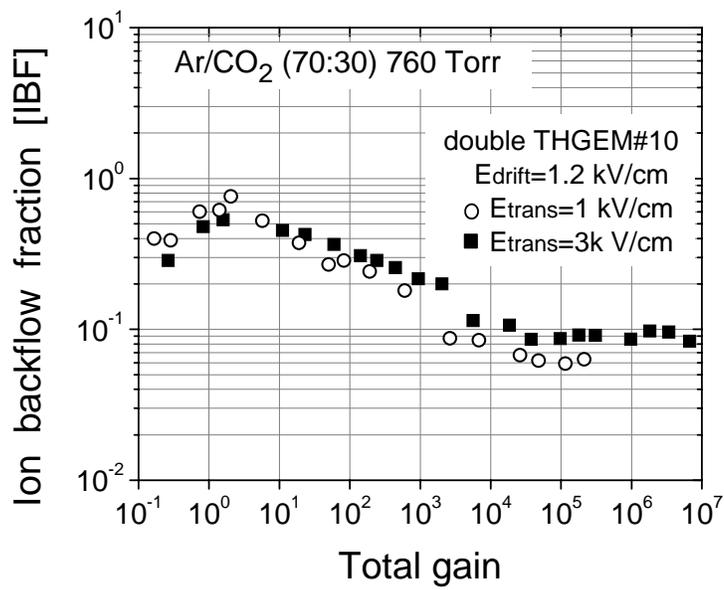

*Fig 26.*

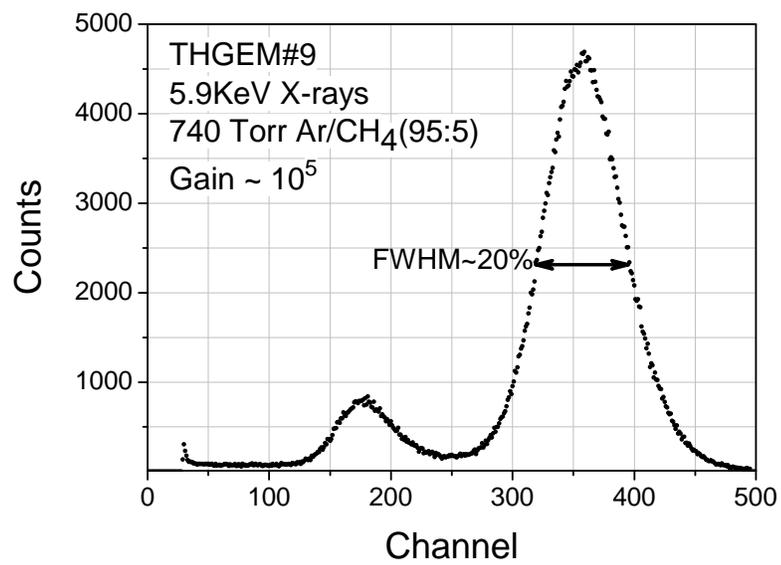

*Figure 27.*